\documentclass[a4paper,12pt]{JHEP3}
\usepackage{subfigure}
\usepackage{longtable}
\usepackage{graphicx}
\usepackage{graphics}
\usepackage{keyval}
\usepackage{trig}
\usepackage{ulem}
\usepackage{graphicx}
\usepackage{dcolumn}
\usepackage{bm}
\def\beq{\begin{equation}}
\def\eeq{\end{equation}}
\def\bey{\begin{eqnarray}}
\def\eey{\end{eqnarray}}
\def\msun{M_\odot}
\def\lsun{L_\odot}

\def\half{{1 \over 2}}

\title{A lower limit on the dark particle mass from dSphs}

\author{G. W. Angus\\
Dipartimento di Fisica Generale ``Amedeo Avogadro", Universit\`a degli Studi di Torino, Via P. Giuria 1, I-10125, Torino, Italy\\
 Istituto Nazionale di Fisica Nucleare (INFN), Sezione di Torino, Torino, Italy\\}

\abstract{We use dwarf spheroidal galaxies as a tool to attempt to put precise lower limits on the mass of the dark matter particle, assuming it is a sterile neutrino. We begin by making cored dark halo fits to the line of sight velocity dispersions as a function of projected radius (taken from Walker et al. 2007) for six of the Milky Way's dwarf spheroidal galaxies. We test Osipkov-Merritt velocity anisotropy profiles, but find that no benefit is gained over constant velocity anisotropy. In contrast to previous attempts, we do not assume any relation between the stellar velocity dispersions and the dark matter ones, but instead we solve directly for the sterile neutrino velocity dispersion at all radii by using the equation of state for a partially degenerate neutrino gas (which ensures hydrostatic equilibrium of the sterile neutrino halo). This yields a 1:1 relation between the sterile neutrino density and velocity dispersion, and therefore gives us an accurate estimate of the Tremaine-Gunn limit at all radii. By varying the sterile neutrino particle mass, we locate the minimum mass for all six dwarf spheroidals such that the Tremaine-Gunn limit is not exceeded at any radius (in particular at the centre). We find sizeable differences between the ranges of feasible sterile neutrino particle mass for each dwarf, but interestingly there exists a small range 270-280eV which is consistent with all dSphs at the 1-$\sigma$ level. }

\begin{document}

\date{\today}
\maketitle

\section{Introduction}
\protect\label{sec:intr}
The lack of success in detecting cold dark matter (CDM) particles \cite{ahmed09} emerging from supersymmetric extensions to the standard model of particle physics (\cite{bergstrom09} and references therein) has led many authors \cite{dodwid,viel05,seljak06,dodster,abazajian06,biermann06,stasielak07,maltoni07,giunti08,melchiorri09,acero09,kusenko09} to investigate the properties of warm dark matter (WDM) candidates such as sterile neutrinos (SNs). WDM might also provide a straight-forward explanation to the missing satellites problem \cite{klypin99,moore99} and the cusp/core problem \cite{mcgaughdeblok,gentile04}, both of which are of particular significance for the dwarf spheroidals of the Milky Way.

Sterile neutrinos are hypothetical neutral leptons, which don't participate in the weak or strong interactions. They can however influence the Universe gravitationally, and are also coupled to the active neutrinos through neutrino oscillations. It is through these oscillations that the most promising mode of unambiguous detection occurs, although no strict limits have been placed on sterile neutrino masses by the controversial experiments LSND \cite{aguilar01} and MiniBoone \cite{maltoni07}.

A thorough study by \cite{boy09} attempted to estimate the minimum mass of SNs in dwarf spheroidals by taking the average stellar velocity dispersions (VDs), assuming the SNs have the same VD and applying the Tremaine-Gunn inequality \cite{tremaine79}, discussed in \S\ref{sec:psc}. In that study, the kinematic data of the dSphs was limited to central velocity dispersions (VDs) of a handful of stars, leaving a degeneracy with the velocity anisotropy. To rectify this general problem, \cite{walker07} published line of sight (los) velocity dispersions of $\sim$6600 stars for 7 dSphs. This large number of stars ($\sim 2400$ for Fornax) means we can bin the stars in increments of the projected radius and reduce the degeneracy with the velocity anisotropy. Therefore the dark matter density profiles of these dSphs are far better constrained, making studies that predate these observations rather obsolete.

There are a further two flaws in certain previous attempts to set limits on the minimum mass of DM particles in dSphs. Firstly, the velocity distribution of the DM particles is often erroneously assumed to be the same as that of the stars. Realistically, the SNs should exist in a stable equilibrium distribution that satisfies hydrostatic equilibrium. In particular, the density of SNs is set by fitting the losVDs of the dSphs, but then the equation of state of a partially degenerate neutrino gas must be used to compute the unique SN VD for a given SN particle mass, $m_{\nu}$. Finally, the phase space density found by combining the necessary SN density and the VD, must be lower than the TG phase space limit. This can make a considerable difference to the deduced minimum mass of SN.

A further flaw is assuming {\it a priori} that the dSph has a cusped dark matter halo. As pointed out by \cite{angusdiaf09} and \cite{evans09}, the losVDs cannot distinguish between cored or cusped halos yet. There is simply too much freedom in the shape of the halo and the velocity anisotropy. The key point here is that cored halos can accomodate lower masses of SN than cuspy halos and it is for this reason, rather than seriously advocating cored halos, that we use them in our analysis. Cored halos were also advocated by \cite{strigari06} and \cite{read06} to help explain the large core size needed in the Fornax dSph to help explain the survival of the globular clusters in orbit under the influence of dynamical friction.

\section{Data and models}
We take the losVDs given in \cite{walker07} and assume that the stellar samples are not polluted by any non-equilibrium stars as suggested in \cite{serra09}. In order to minimise the different density profiles used in our modelling we apply the same cored density models for the DM halos to give new fits to the 3-D light profiles of the dSphs which are excellent representations of King models interior to the classical tidal radius (see Fig~\ref{fig:c1}. Beyond the tidal radius, the King models crash towards zero density, but our cored models decline with a constant logarithmic slope - which might be more in tune with recent observations of the edges of dSphs (e.g. \cite{munoz07}). The density models are defined as follows

\beq
\protect\label{eqn:angd}
\rho_i(r)=\rho_{o,i} \left[1+\left({r \over r_i}\right)^2 \right]^{-\gamma_i},
\eeq
where the core radius, central density and asymptotic slope are given by $r_i$, $\rho_{o,i}$ and $\gamma_i$ respectively. The subscript ``i'' corresponds to a ``$*$'' for the stellar model or a ``$\nu$'' for the neutrinos. The details for the new stellar density fits are given in Table~1, as are the fitted dark halo parameters discussed later.

\section{Solving for sterile neutrino equilibrium configurations}
\subsection{dSph Jeans analysis} 
The density of SNs in dSphs is given by the following steps. Firstly, we must take the spherical Jeans equations
\beq
\protect\label{eqn:jeans}
{d \over dr}\sigma_r^2(r) + \left[\alpha(r)+2\beta(r) \right]r^{-1} \sigma_r^2(r) = -g(r)
\eeq
which solves for the radial velocity dispersion, $\sigma_r^2(r)$. For the velocity anisotropy, we initially tried an Osipkov-Merritt (OM) model
\beq
\protect\label{eqn:beta}
\beta(r)=\beta_0+\beta_{\infty}{1 \over 1 + \left({r_{\beta} \over r}\right)^2},
\eeq
where $\beta_{0}$, $\beta_{\infty}$ and $r_{\beta}$ are free parameters (although -$\infty<\beta<1$). However, we found no benefit from this model over constant anisotropy, although we still solve the equations for the OM model.

The function $\alpha(r)$ is the logarithmic derivative of the 3-D light profile defined in Eq~\ref{eqn:angd} such that

\beq
\protect\label{eqn:alpha}
\alpha(r)={r \over \rho_*(r)} {d\rho_* \over dr}={-2\gamma_* \over 1+ ({r_* \over r})^2}.
\eeq
The density model for the SN halo, given by Eq~\ref{eqn:angd} also defines the gravity at all radii (where we ignore the puny contribution from $\rho_*$)

\beq
\protect\label{eqn:grav}
g(r)={4\pi G \over r^2}\int_0^r \rho_{\nu}(\hat{r})\hat{r}^2 d\hat{r}.
\eeq
From here we use the integrating factor
\beq
\protect\label{eqn:grav}
I(r)=(r^2+r_*^2)^{-\gamma_*}(r^2+r_{\beta}^2)^{\beta_{\infty}}r^{2\beta_0},
\eeq
which finally allows us to numerically integrate to find the radial velocity dispersions

\beq
\protect\label{eqn:rhonu1}
\sigma_r^2(r)=I^{-1}(r)\int^{\infty}_r I(\hat{r})g(\hat{r})d\hat{r}.
\eeq

We then cast the radial velocity dispersion into the line of sight to compare with the observations, and minimise the $\chi^2$. The best fit SN density parameters, $\rho_{o,\nu}$, $r_{\nu}$  and velocity anisotropy are given in Table~1, where in order to minimise the freedom of our models, we force $\gamma_{\nu}=1.5$. The error bars on the 3 free parameters correspond to the differences the parameters would need to have from the best fit in order to raise the $\chi^2$ by 3.5. As expected, this is a large parameter space, because changes in the combination of $\rho_{o,\nu}$ and $r_{\nu}$ can be counter balanced by the variable velocity anisotropy. Fortunately, the shape of the losVD profile keeps these combinations finite. Furthermore, although the individual ranges in $\rho_{o,\nu}$ and $r_{\nu}$ are vast, combined measures like the enclosed masses (for instance $M_{300}$ in Table~1) are much less extensive. In the middle panel of Fig~1, the best fit model is given by the solid line and the dashed lines are the 1-$\sigma$ errors on the velocity anisotropy.

\subsection{Sterile neutrino hydrostatic equilibrium}
Now we have a density of SNs in each dSph fixed by the stellar losVDs: at least it is the lowest density exactly required for hydrostatic equilibrium of the {\it stars} in the dSph, that has centrally isotropic orbits and fits the data well. In addition to this constraint, there is an equation for hydrostatic equilibrium of the SNs themselves. This invokes the equation of state of neutrinos and gives us a relation that defines the density and pressure of a partially degenerate neutrino gas, coupled via the hydrostatic equilibrium relation ${d \over dr}P_{\nu} = -\rho_{\nu}g(r)$.

The exact procedure to find the velocity dispersion required for hydrostatic equilibrium of the SN halo is rather lengthy, but is given in its full glory in \cite{afd}. The end result is that there is a one to one relation between the SN density, $\rho_{\nu}$, and gravity (which is defined by $\rho_{\nu}$ in Eq~\ref{eqn:grav}) and the SN velocity dispersion, with the SN particle mass the only tunable parameter.

\subsection{Phase space constraints}
\protect\label{sec:psc}
We now have a unique correlation between the density of SNs and their velocity dispersions. This is important because these two variables define the phase space distribution of the SNs, to which there is a fundamental limit.

Liouville's theorem states that (in the absence of encounters) flow in phase space is incompressible and that each element of phase-space density is conserved along the flow lines. However, this only applies to the fine-grained phase space density of an infinitesimal region. Rather, for the observable, which is the coarse-grained (macroscopic) phase-space density, we simply must not exceed the maximum of the fine-grained one.

Thus, the SN phase-space density must not increase during collapse, from its starting value of $\half g_{\nu}h^{-3}m_{\nu}^{4}$ (which is half the Pauli limit), to its current maximum value of $\rho_{\nu}(r)[2\pi\sigma_{\nu}^2(r)]^{-1.5}$ (where we assume the velocity distribution is a Gaussian with dispersion $\sigma_{\nu}$). This limit is called the Tremaine-Gunn (TG) limit and rearranging it in terms of the critical density for a given SN particle mass gives us

\beq
\protect\label{eqn:tg}
\rho_{\nu_s,TG}(r)=215\left({m_{\nu} \over 1~eV/c^2}\right)^4\left({\sigma_{\nu}(r) \over c} \right)^3\msun pc^{-3}.
\eeq
The important thing to bear in mind here is that this is a more accurate value of the TG limit because it is calculated from the unique, derived SN VD exactly necessary for hydrostatic equilibrium and not by assuming some relation between the stellar VD and the sterile neutrino VD, as is often the case \cite{gentile08,angus08,boy09}.  The only assumption is that the velocity distributions are locally Maxwellian everywhere.

This defines the TG limit as a function of radius, since we know the exact value of the VD at all radii from solving for the SN hydrostatic equilibrium. Our final step is to vary the mass of the SN such that the TG limit is reached exactly at the centre. For the best fit parameters, the minimum SN mass is given in Table~1 and the errors are propogated by finding the minimum SN mass needed when the extreme values of $\rho_{o,\nu}$ and $r_{\nu}$ are used ($\beta$ plays no part).

\section{Results}

\begin{figure*}
\def\subfigtopskip{0pt} 
\def\subfigbottomskip{4pt}
\def\subfigcapskip{1pt}
\centering
\begin{tabular}{c}
\subfigure{
\includegraphics[angle=0,width=16.0cm]{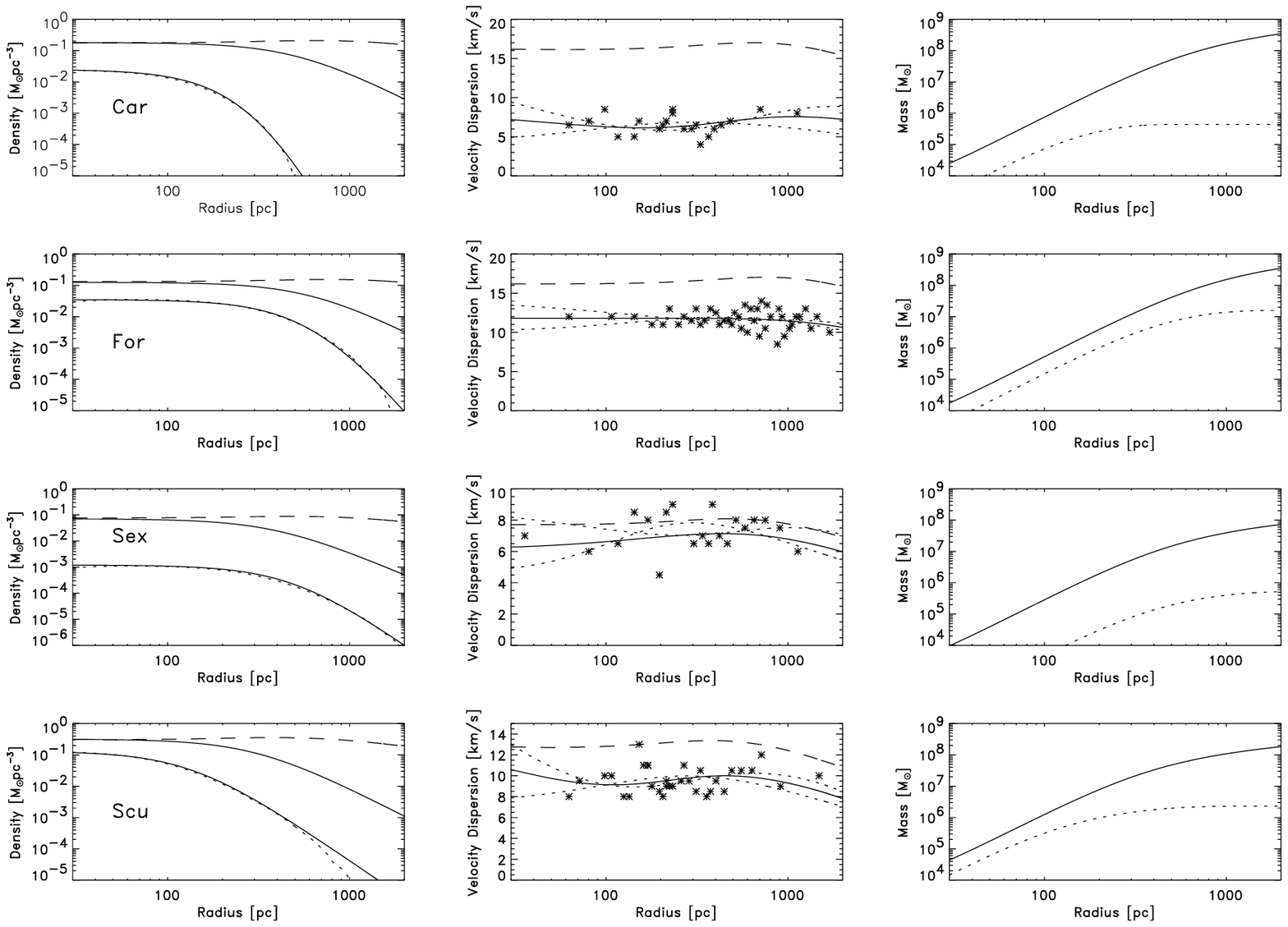}
}\\
\subfigure{
\includegraphics[angle=0,width=16.0cm]{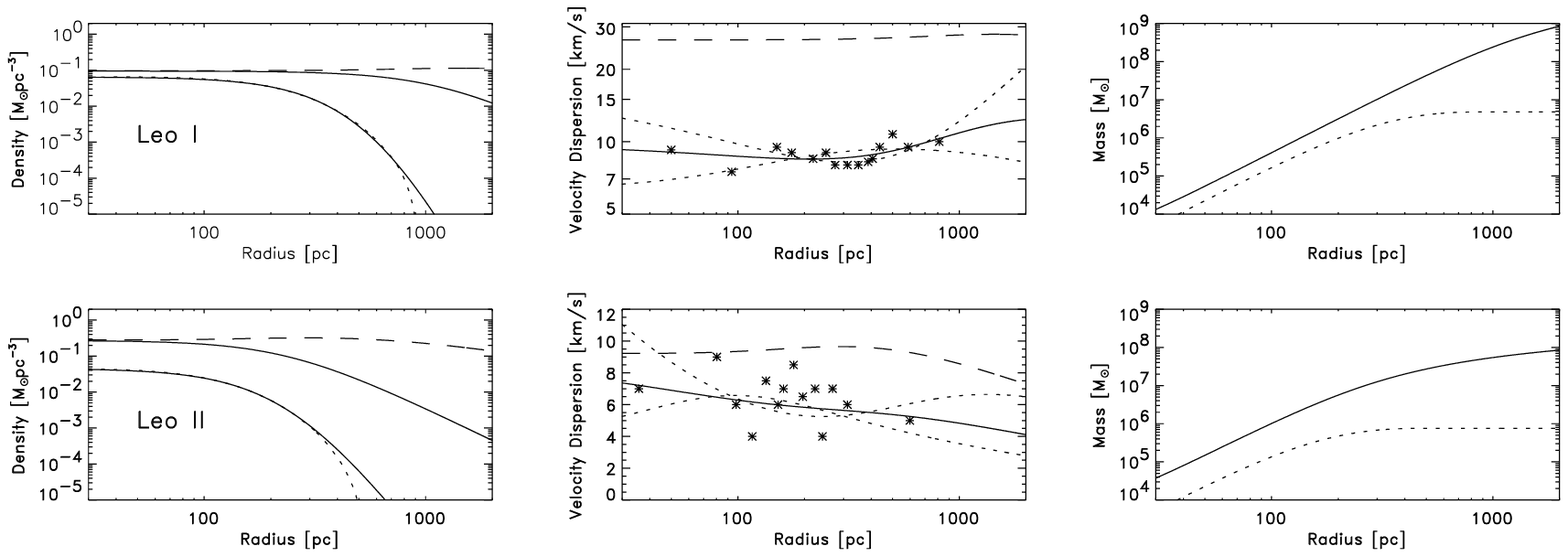}
}\\
\end{tabular}
\vskip -6cm
\caption{Each row refers to a specific dwarf spheroidal (identified in the left hand panel), whereas the columns are properties of that dwarf spheroidal. In the left hand panels we plot the 3-D density against radius for the best fit sterile neutrino halo (solid), and the corresponding TG limit (dashed). In addition we plot the King model stellar density (dotted, with $M/L$=1) and overplot the best fit cored density model of Eq~1. In the central panels we plot the inferred 3-D velocity dispersion of the sterile neutrinos (dashed) required for their hydrostatic equilibrium, and the projected stellar line of sight velocity dispersions (solid) - which are fitted to the data points (asterix symbols) from Walker et al. (2007). The dotted lines are the upper and lower 1-$\sigma$ errors given by the extremes of velocity anisotropy. In the right hand panels we plot enclosed sterile neutrino halo (solid) and stellar (dotted) masses against radius.}
\label{fig:c1}
\end{figure*}

In Fig~\ref{fig:c1} we present the densities, VDs and enclosed mass profiles for the dark halos and stars for the six dSphs. In the left hand panels we plot the SN density (solid line type) and the TG limit of the dSph (dashed) against radius. In addition to this, we show the King model densities (with unity $M/L$) along with the Eq~1 density models (with parameters given in Table 1). In the middle panels we plot both the derived SN (dotted) VD and stellar losVD (solid) which is fitted to the data points taken from \cite{walker07}. In the right hand panels, we plot each dSph's enclosed SN dark halo (solid) and stellar (dashed) masses.

It is encouraging that the velocity dispersions of the SNs are similar to the losVDs of the stars, and that they are smooth. It is worth mentioning here that the equilibrium SN velocity dispersion (and hence TG limit) depends critically on the slope the SN density profile. As can be seen in the left hand panels of Fig~\ref{fig:c1}, the TG limit remains fairly constant at all radii. Our SN halos, although cored in name, continue to increase slowly towards the centre in all cases. It is this behaviour that forces the TG limit to fall slightly (a factor of 2 or so only in these cases) from slightly outside the centre. If the density increases relatively rapidly, the phase space density increases towards the TG limit (also towards the Fermi limit). As it does so, the classical pressure required for hydrostatic equilibrium ($P_{\nu,c} \propto  \rho_{\nu}\sigma_{\nu}^2$) is gradually replaced by Fermi pressure ($P_{\nu,f} \propto \rho_{\nu}^{5/3}$) causing the VD to decrease, and hence the TG limit to decrease. Since we cannot decrease the SN density any further (without sacrificing our fits to the stellar losVDs), if the SN particle mass is too small, then the TG limit will dip below the required density and will be unfeasible. 

This can be thought of another way: the more cuspy the SN density needs to be, the larger the SN particle mass must be to evade the TG limit. It is by increasing the SN particle mass, that we increase the TG limit at all radii and choose the mass that causes the TG limit to drop to the required density only at the centre.

The minimum SN masses required for our six dSphs are given in Table 1 and are plotted against dSph luminosity in Fig~2. Interestingly, although the two Leo dSphs provide the extremes of the preferred particle mass ($160^{+130}_{-115}$ and $460^{+470}_{-240}$ for Leo I and II respectively), the other four lie in a relatively tight range between 250 and 380~$eV$. Surprisingly, there exists a tiny 10~$eV$ interval (270-280~$eV$) which all six dSphs are consistent with at 1-$\sigma$. This minimum SN particle mass is significantly smaller, in all but the case of Sextans, than the traditional estimates of the TG limit (given in Table 1) found using Eq~2 of \cite{madsen91}.

Our approach also improves the limits on the lowest allowed SN mass because we do not need to assume a relation between the stellar losVD and the velocity dispersion of the SNs (we solve directly for it). In five of the six cases, the SN VD is 20-100\% larger than the stellar losVD, leaving the unusual case of Leo I which prefers a very large core - meaning a lower mass SN with a larger velocity dispersion can be accomodated. Furthermore, since we use cored neutrino halo models, we allow less massive neutrinos than cuspier halos would allow.

\cite{strigari08} proposed the common mass scale where each dSph dark halo should have $\sim 10^7\msun$ within 300~pc, the ultra-faint dSphs \cite{willman06,zucker06,belokurov06,belokurov07,simon07} discovered by the SDSS should pose a more stringent test for us. However, their small samples of stars mean we cannot reliably use our method to determine an equilibrium configuration for the SNs. In addition, questions have been raised by \cite{nied09} and \cite{serra09} about how well interlopers have been dealt with in these systems. Therefore, our limits should be the strongest and most reliable ones possible. 

On the same topic, our cored halo models offer a sanity check on the cuspier halo models used to constrain the mass within 300~pc ($M_{300}$) for all dSphs by \cite{strigari08}. Like \cite{strigari08}, our $M_{300}$s do not appear to be correlated with satellite luminosity, although the preferred common mass scale is very slightly (10-20\%) higher.

\begin{figure*}
\def\subfigtopskip{0pt} 
\def\subfigbottomskip{4pt}
\def\subfigcapskip{1pt}
\centering
\begin{tabular}{cc}
\subfigure{
\includegraphics[angle=0,width=8.0cm]{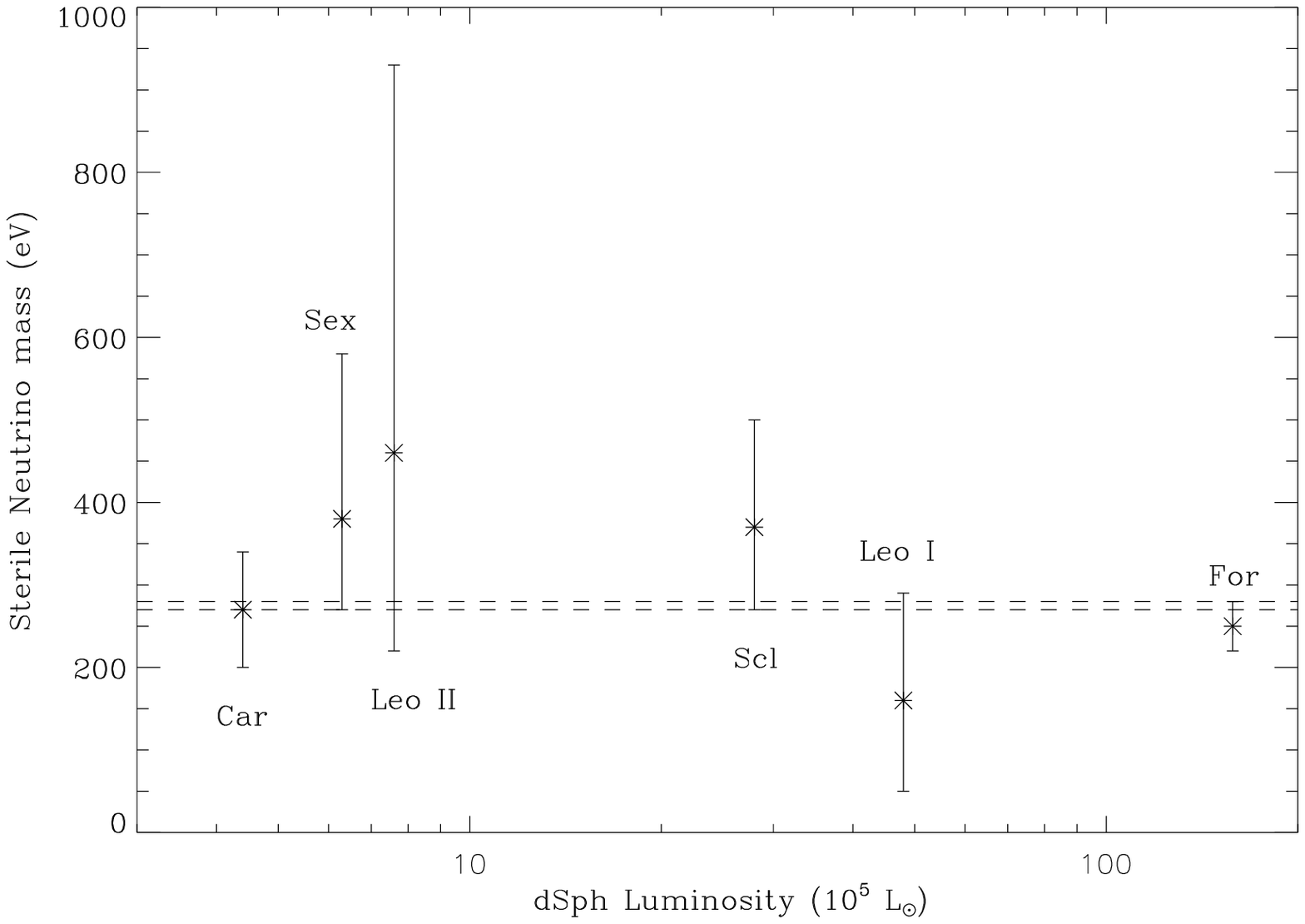}
}
\subfigure{
\includegraphics[angle=0,width=8.0cm]{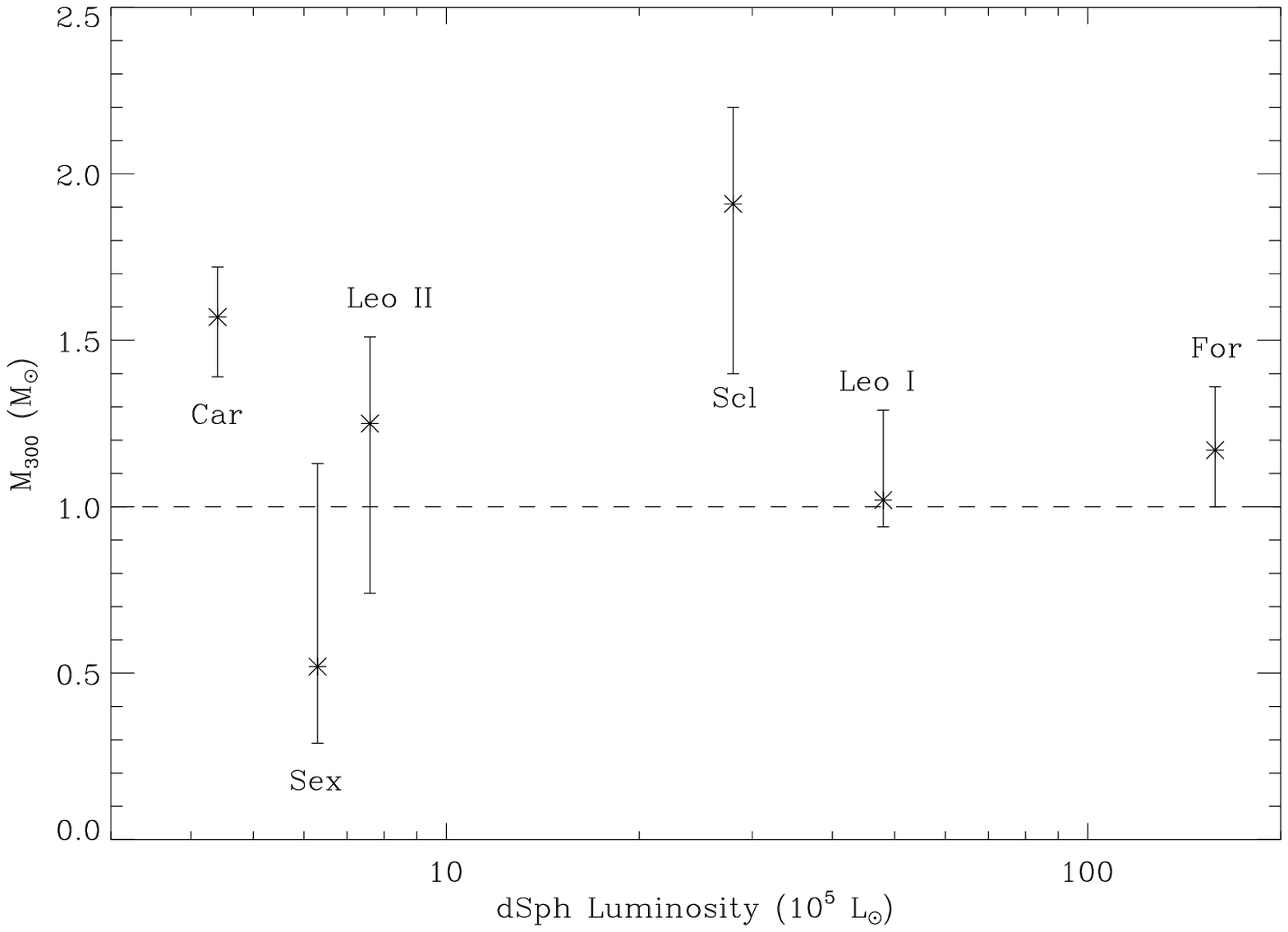}
}\\
\end{tabular}
\caption{The left hand figure shows the minumum sterile neutrino masses plotted against their luminosities. The thin dashed strip shows the range of neutrino masses that are consistent with all six dSphs at the 1-$\sigma$ level. In the right hand figure, we plot $M_{300}$ against luminosity. Our six dSphs seem to prefer a slightly higher value of $M_{300}$ than the value deduced in Strigari et al. (2008), perhaps only due to the different dark matter density profiles and velocity anisotropies.}
\label{fig:c3}
\end{figure*}
\begin{table*}
\tiny
\begin{tabular}{cccccccccccccccc}
dSph    & $R_{MW}$   & $L_v$ & $\rho_{o,*}$  & $r_*$ & $\gamma_*$& $\rho_{o,\nu}$ & $r_{\nu}$ & $\beta$& $\chi^2$/n& $M_{300}$  &Old $m_{\nu}$ &Our $m_{\nu}$\\ 
	&   kpc      &$10^5\lsun$&$10^{-3}\lsun pc^{-3}$&pc& &$\msun pc^{-3}$&pc&& &$10^7\msun $&$eV$&$eV$\\
Carina  & 101$\pm$5  & 4.4 $\pm$ 1.1       & 25  & 310  & 5.5  &$0.18^{+0.088}_{-0.01}$&$518^{+299}_{-255}$ &$0.3^{+0.35}_{-0.9}$ & 13.7/22&$1.57^{+0.15}_{-0.18}$& $470^{+80}_{-55}$ &$270^{+150}_{-70}$ \\
Fornax  & 138$\pm$8  & 158 $\pm$ 16  & 35  & 650  & 3.5 &$0.125^{+0.031}_{-0.024}$  &$622^{+123}_{-106}$&$0.0^{+0.2}_{-0.25}$  &58.6/43&$1.17^{+0.19}_{-0.17}$&$280^{+40}_{-30}$ &$250^{+30}_{-30}$\\
Sculptor& 87$\pm$4   & 28 $\pm$ 4 & 130 & 135 & 2.0  & $0.316^{+0.411}_{-0.155}$  &$307^{+209}_{-131}$ &$0.5^{+0.35}_{-0.8}$ &23.3/33 &$1.91^{+0.29}_{-0.51}$&$565^{+105}_{-70}$&$370^{+130}_{-100}$\\
Sextans & 95$\pm$4   & 6.3 $\pm$ 1.4      & 1.2 & 500   & 2.5  & $0.07^{+0.37}_{-0.04}$  &$394^{+321}_{-235}$&$-0.2^{+0.5}_{-1.8}$  &23.3/20&$0.52^{+0.61}_{-0.23}$& $350^{+70}_{-45}$ &$380^{+200}_{-110}$\\
Leo I   & 257$\pm$8  & 48 $\pm$ 5    & 65  & 600   & 6.0  & $0.095^{+0.055}_{-0.012}$  &$1172^{+6300}_{-672}$  &$0.17^{+0.33}_{-0.67}$   &3.3/15&$1.02^{+0.27}_{-0.08}$& $450^{+70}_{-50}$ &$160^{+130}_{-115}$\\
Leo II  & 233$\pm$15 & 7.6 $\pm$ 3   & 45  & 245 & 4.0  & $0.271^{+2.33}_{-0.126}$ &$239^{+465}_{-185}$& $0.3^{+0.6}_{-2.3}$&14.1/15 &$1.25^{+0.26}_{-0.51}$& $490^{+80}_{-55}$&$460^{+470}_{-240}$\\
\end{tabular}
\caption{\tiny Here we list all the parameters and results pertaining to the fits of our cored sterile neutrino halos to the six dSphs and the best fit light profiles for each dSph. Also given are the sterile neutrino mass within 300~pc and the minimum mass of sterile neutrino particle that we calculate can exist in hydrostatic equilibrium. These are compared to the traditional Tremaine-Gunn limits (Madsen 1991). In each case, the velocity anisotropies are constant. The distances and luminosities of the dSphs are all taken from Table 1 of Angus (2008), where the original references are given. Error bars on the three free fit parameters are at the 1-$\sigma$ error and these propogate simply into the minimum neutrino mass and $M_{300}$.}
\end{table*}
\section{Summary and Conclusion}
We have taken the data of \cite{walker07} on the line of sight velocity dispersions of six dwarf spheroidal galaxies (dSphs) and fitted cored sterile neutrino halos assuming some anisotropic orbits for the stars. From this we solved for the velocity dispersion of the dark halos at all radii by solving the equation of state for a partially degenerate neutrino gas. This allowed us to infer the lower limit on the dark matter particle mass (assuming it is a sterile neutrino) and found it is in the range 160-460~$eV$. This limit is lower than estimates achieved with the traditional Tremaine-Gunn approach and intriguingly, there exists a range of minimum sterile neutrino mass (270-280~$eV$) which is consistent with all six dSphs at the 1-$\sigma$ level.

\section{Acknowledgments} GWA's research is supported by the University of Torino and Regione Piemonte. Partial support from the INFN grant PD51 is also gratefully acknowledged as are the insightful comments of the anonymous referee.

\bibliography{sndsph}
\end{document}